\documentstyle[preprint,aps,epsf,eqsecnum]{revtex}
\begin{document}
\preprint{UGVA-DPT 1995/05-890 ;
BARI-TH/95-200}
\title{STRONG COUPLING OF EXCITED HEAVY MESONS\footnote{Partially
supported by the Swiss National Foundation}}
\author{P. Colangelo, F. De Fazio and G. Nardulli}
\address{Dipartimento di Fisica, Univ.
di Bari\\
I.N.F.N., Sezione di Bari}
\author{N. Di Bartolomeo and R. Gatto}
\address{D\'epartement de Physique Th\'eorique, Univ. de Gen\`eve}

\date{May 1995}
\maketitle

\begin{abstract}
We compute the strong coupling constant $G_{B^{**} B \pi} \; (G_{D^{**} D
\pi})$, where
$B^{**}$ ($D^{**}$) is the $0^+$ $P-$wave $b \bar q \; (c \bar q)$ state,
by QCD sum rules and by
light-cone sum rules. The two methods give compatible results in the limit
$m_Q \to \infty$, with a rather large value of the coupling constant.
We apply the results to the calculation of the hadronic
widths of the positive parity $B$ and $D$ states
and to the chiral loop contribution to
the ratio $f_{D_s}/f_D$.
\end{abstract}

\section{Introduction}

In the light quark ($q=u,d,s$) zero mass limit ($m_q \to 0$) and in the
heavy quark ($Q=c,b$) infinite mass limit ($m_Q \to \infty$) Quantum
Chromodynamics exhibits symmetries that are not present in the finite
mass theory:  chiral $SU(3)_L \times SU(3)_R$ symmetry, heavy quark
spin and flavour symmetries
\cite{hqet}, as well as the velocity superselection rule \cite{georgi},
 valid insofar only strong interactions are considered.

All these symmetries can be used to build up an effective chiral lagrangian
\cite{chirlag} for light pseudoscalar mesons and heavy $(Q\bar q)$ negative
parity $(J^P\,=\,0^-,\, 1^-)$ mesons. This effective lagrangian
contains a symmetric term plus corrections to the heavy-light symmetries
such as, for example, terms proportional to  powers of $1/m_Q$
or terms proportional to powers of the light quark masses $m_q$. In the spirit
of the chiral effective theory, the resulting lagrangian is also an expansion
in the light meson fields derivatives.

Besides the octet of the pseudo-Goldstone bosons $\pi, K, \eta$ and the
$(D,\, D^*)$, $(B,\, B^*)$ states,
one can include the light vector mesons belonging to
the low lying $SU(3)$ nonet: $\rho , \, K^* , \, \omega_8 , \, \omega_0$,
using the so-called hidden symmetry approach \cite{casalbuo,schec}.
As well
known, in this approach \cite{bando} the lagrangian exhibits an extra local
gauge symmetry $SU(3)_H$, and the $1^-$ light meson octet represents its
gauge bosons. They acquire a mass because  $SU(3)_H$ is spontaneously
broken. Quite recently it has been observed \cite{georgi2} that, if  $SU(3)_H$
is unbroken, a new symmetry (vector symmetry) arises. Its implications for
the heavy-light chiral lagrangian have been examined in Ref. \cite{mannel}.
Additional notions arise, in general, when also axial-vector bosons are
present \cite{bessav}.

Another extension of the heavy-light lagrangian is obtained by
including effective fields describing  positive parity
$(Q\bar q)$ mesons. According to the value of the angular momentum of the light
degrees of freedom,
$(s^P_\ell \, = \, {\frac{1}{2}}^+ \;, \; {\frac{3}{2}}^+)$
the Heavy Quark Effective Theory \cite{isgur,falkluke}
predicts the existence of two multiplets, the first one comprising
$0^+$ and $1^+$ mesons, the second one containing $1^+ , \; 2^+$ states.
The role of the ${\frac{1}{2}}^+$ doublet  $(0^+  , \; 1^+)$ in some
applications of chiral perturbation theory has been considered in
\cite{falk}\footnote{There is no coupling of the
$(s^P_\ell \, = \; {\frac{3}{2}}^+)$ states to the $(0^-  , \; 1^-)$
heavy meson doublet and to the pseudoscalar Goldstone bosons at the lowest
order in the chiral expansion \cite{falkluke}.}. Another application
is in the realm of the semileptonic $D$ and $B$ decays \cite{casalbuo}.

The aim of the present paper is to give an estimate of the
strong coupling constant describing the interaction of the pseudoscalar light
mesons with the
positive parity $(s^P_\ell \, = \; {\frac{1}{2}}^+)$ and negative
parity $(s^P_\ell \, = \; {\frac{1}{2}}^-)$ heavy mesons and to make
quantitative estimates of the effect of
the positive parity heavy mesons in some calculations in chiral
perturbation theory.

After a review of the heavy-light chiral
lagrangian in Section II, we consider two sum rules for these
coupling constants: the first one, based on the method of the single
Borel transform (in the soft pion limit), is
discussed in  Section III, while in Section IV we
derive these couplings by the method of the
light-cone sum rules \cite{chern1} (for a review see
\cite{chern2}).

In Section V we compute the decay widths
of the excited positive parity heavy mesons, and we estimate the
mixing angle between $\frac{1}{2}^+$ and $\frac{3}{2}^+$ axial vector states.

In Section VI we comment on the role of the
positive parity heavy mesons
in the chiral loop contributions to the ratio $f_{D_s}/f_D$, which,
as observed in \cite{falk}, may be considerable. We
find the numerical result:
\begin{equation}
\frac{f_{D_s}}{f_D} \; = \; 1.09
\end{equation}
where part of the SU(3) violation effect $f_{D_s}/f_D \neq 1$
may be attributed to positive parity states.

Finally, in Section VII  we draw our conclusions.
\section{The heavy-light chiral lagrangian}
In the effective heavy-light chiral lagrangian, the ground state
$(s^P_\ell \, = \; {\frac{1}{2}}^-)$ heavy mesons  are described by the
 $4 \times 4$ Dirac matrix
\begin{equation}
H_a = \frac{(1+{\rlap{v}/})}{2}[P_{a\mu}^*\gamma^\mu-P_a\gamma_5]
\label{h}
\end{equation}
where $v$ is the heavy meson velocity,
$P^{*\mu}_a$ and $P_a$ are annihilation operators
of the $1^-$ and $0^-$ $Q{\bar q}_a$ mesons
($a=1,2,3$ for $u,d$ and $s$): for charm, they are $D^*$ and $D$
respectively.
Similarly, the positive parity $1^+$ and $0^+$
$(s^P_\ell \, = \; {\frac{1}{2}}^+)$
are described by
\begin{equation}
S_a=\frac{1+{\rlap{v}/}}{2} \left[D_1^\mu\gamma_\mu\gamma_5-D_0\right] \;.
\label{s}
\end{equation}

It should be observed that all the operators appearing in Eq. (\ref{h}) and
Eq. (\ref{s}) have dimension $\frac{3}{2}$ since they contain a factor
${\sqrt {m_P}}$
in their definition.

As for the octet of the pseudo Goldstone bosons, one uses the exponential form:
\begin{equation}
\xi=\exp{\frac{iM}{f_{\pi}}}
\end{equation}
where
\begin{equation}
{M}=
\left (\begin{array}{ccc}
\sqrt{\frac{1}{2}}\pi^0+\sqrt{\frac{1}{6}}\eta & \pi^+ & K^+\nonumber\\
\pi^- & -\sqrt{\frac{1}{2}}\pi^0+\sqrt{\frac{1}{6}}\eta & K^0\\
K^- & {\bar K}^0 &-\sqrt{\frac{2}{3}}\eta
\end{array}\right )
\end{equation}
and $f_{\pi}=132 \; MeV$.

The lagrangian describing the fields $H$, $S$ and $\xi$ and their interactions,
under the hypothesis of chiral and spin-flavour symmetry and at the lowest
order in light mesons derivatives is:
\begin{eqnarray}
{\cal L} &=& \frac{f_{\pi}^2}{8}<\partial^\mu\Sigma\partial_\mu
\Sigma^\dagger > +i < H_b v^\mu D_{\mu ba} {\bar H}_a >  \nonumber\\
& + & < S_b \;( i \; v^\mu D_{\mu ba} \; - \; \delta_{ba} \; \Delta)
{\bar S}_a > +\; i \, g <H_b \gamma_\mu \gamma_5 {\cal A}^\mu_{ba}
{\bar H}_a> \nonumber\\
& + & i \, g' <S_b \gamma_\mu \gamma_5 {\cal A}^\mu_{ba} {\bar S}_a>
+ \,[ i \, h <S_b \gamma_\mu \gamma_5 {\cal A}^\mu_{ba} {\bar H}_a> \;
+ \; h.c.] \label{L}
\end{eqnarray}
where $<\ldots >$ means the trace, and
\begin{eqnarray}
D_{\mu ba}&=&\delta_{ba}\partial_\mu+{\cal V}_{\mu ba}
=\delta_{ba}\partial_\mu+\frac{1}{2}\left(\xi^\dagger\partial_\mu \xi
+\xi\partial_\mu \xi^\dagger\right)_{ba}\\
{\cal A}_{\mu ba}&=&\frac{1}{2}\left(\xi^\dagger\partial_\mu \xi-\xi
\partial_\mu \xi^\dagger\right)_{ba} \; ;
\end{eqnarray}
$\Sigma= \xi^2$ and $\Delta$ is the mass splitting of the $S_a$ states from
the ground state $H_a$. Numerically we use $\Delta\,=\, 500 \pm 100$ MeV,
an estimate based on quark model \cite{isgur2} and QCD sum rules \cite{noi}
computations of the masses of the excited
$s_{\ell}^P \, = \,
{\frac{1}{2}}^+$ mesons.

In Ref. \cite{g} a QCD sum rule has been considered to compute the strong
coupling constant $g$ appearing in Eq.(\ref{L}), with the result
(at the order $\alpha_s =\, 0$):
\begin{equation}
g = 0.39 \pm 0.16 \; .
\end{equation}
This result, valid in the soft pion limit, has been confirmed by a
subsequent, independent analysis, based on the method of light-cone sum rules
\cite{ruckl}:
\begin{equation}
g = 0.32 \pm 0.02 \; .
\end{equation}
The results of similar sum rules for the strong coupling constant
$h$ in Eq.(\ref{L}) will be presented in the subsequent Sections.
\section{QCD sum rule for \lowercase{$h$}}

Let use define
the strong amplitude
\begin{equation}
 G_{B^{**} B \pi} \; = \; <\pi^+(q)~B^o(q_2) | B^{**+}(q_1)>
\label{g}
\end{equation}
where $B^{**}$ is the $0^+$ state in the
$s_{\ell}^P = {\frac{1}{2}^+}$ doublet.
The amplitude $G_{B^{**} B \pi}$ is related to
the strong coupling constant $h$ appearing in the heavy-light chiral lagrangian
(\ref{L}) by the formula:
\begin{equation}
G_{B^{**} B \pi} \;= - \; {\sqrt {m_B m_{B^{**}}}} \; \;
{{ m^2_{B^{**}}- m^2_B}\over{m_{B^{**}}}}\;
{{h}\over{f_\pi}} \; . \label{grel}
\end{equation}

In the  limit $m_b \to \infty$ one has:
\begin{eqnarray}
m_B & = & m_b + \omega + {\cal O} \left( {1\over{m_b}}\right) \nonumber \\
m_{B^{**}} - m_B & = &  \Delta \; + \; {\cal O} \left( {1\over{m_b}} \right)
\label{para}
\end{eqnarray}
and
\begin{equation}
G_{B^{**} B \pi} \simeq  -  \frac{2 h}{f_{\pi}} \; m_b \; \Delta  \label{gas}
\end{equation}
In order to derive a sum rule for $G_{B^{**} B \pi}$ and $h$,
we consider the correlator:
\begin{equation}
A_\mu= i \int dx <\pi^+(q)| T( j_5(x) V_\mu(0) |0> e^{-i q_2 x} =
A(q_1^2, q_2^2, q^2) q_\mu + B(q_1^2, q_2^2, q^2) P_\mu
\label{corr1}
\end{equation}
where $j_5={\overline u} i \gamma_5 b$, $V_{\mu}={\overline b}
\gamma_{\mu} d$, $ q= q_1 - q_2$ and $P= q_1 + q_2$. The
scalar functions $A$  and $B$ satisfy dispersion relations (D.R.)
that are computed, according to
the method of QCD sum rules, in two ways: either by saturating the dispersion
relation by physical hadronic states, or by means of the operator product
expansion (O.P.E.).  In Ref \cite{g} we considered the dispersion relation
for the scalar function
$A$ and we used it to compute the coupling constant
$G_{B^{*} B \pi}$ defined by the matrix element:
\begin{equation}
<\pi^+(q)~B^o(q_2) | B^{*+}(q_1,\epsilon)> \;= \;
G_{B^{*} B \pi} \; \epsilon_\mu \cdot q^\mu\; .
\label{gbb}
\end{equation}
We obtained the result:
\begin{equation}
f_B \; f_{B^*} \; G_{B^{*} B \pi} = 0.56 \pm 0.12 \; GeV^2 \label{gb}
\end{equation}
while for  charm  we obtained:
\begin{equation}
f_D \; f_{D^*} \; G_{D^{*} D \pi} = 0.34 \pm 0.08 \; GeV^2 \; . \label{gd}
\end{equation}
The leptonic decay constants appearing in the previous equations are
defined as follows:
\begin{eqnarray}
<0| {\overline b}\gamma^\mu \gamma_5 d|B(p)> &=& i f_B p^\mu \nonumber \\
<0| {\overline u} \gamma^\mu b|B^{*}(p, \epsilon)> &=&
i \; m_{B^*} f_{B^{*}} \; \epsilon^\mu \; .
\end{eqnarray}
It should be noticed that the phases in the previous equation are consistent
with the definition of the weak current in the effective theory, see below
Eq.(\ref{wcurr}).

In the following we shall make use also of the
matrix element:
\begin{equation}
<0| {\overline u} \gamma^\mu b|{B^{**}}(p)> =  i f_{B^{**}} \; p^\mu \; .
\end{equation}
It may be observed that the
hadronic side of the sum rule for $A$ also includes the $B^{**}$
pole in the variable $q_1^2$ (the low lying pole in the variable
$q_2^2$  is provided, of course, by the $B$ meson). In \cite{g}
one gets rid of it by exploiting the tiny mass difference between $B$ and
$B^*$, which allows to include the $J^P=0^+$ pole in the so called
{\it parasitic terms} that may be appropriately parametrized. Here we
are interested precisely
in this pole and, in order  to obtain it,  we consider the dispersion
relation for the scalar function $B$ in Eq.(\ref{corr1}) that should
be written in general as follows \cite{eletski} :
\begin{eqnarray}
B(q_1^2,q_2^2,q^2)&=&
{1\over{\pi^2}} \int ds ds' {{\rho(s,s',q^2)}\over{(s-q_1^2)
(s'-q_2^2)}} +  \nonumber\\
 &+& P_1(q_1^2) \int ds' {{\rho_1(s',q^2 )}\over{s' -q_2^2}} +
P_2(q_2^2) \int ds {{\rho_2(s,q^2)}\over{s-q_1^2}} + \,P_3(q_1^2,q_2^2,q^2) \;.
\nonumber\\ \label{had}
\end{eqnarray}
As proven in
\cite{smilga1} $P_1(q_1^2)\;=\;P_2(q_2^2)\;=0$ since only this
value of the polynomials  $P_1(q_1^2)$ and $P_2(q_2^2)$ is compatible
with the vanishing of the form factors for large values of their
arguments, as predicted by quark counting rules.
The
 subtraction polynomial $P_3(q_1^2,q_2^2)$ does not contribute to the
sum rule since it vanishes  after the Borel transform; therefore
we shall neglect it in the sequel.

In order to compute the D.R. for $B$,
in this Section we make the approximation
of the soft pion limit (S.P.L.): $ q \to 0$. This approximation presents the
advantage of a considerable simplification of the calculations; moreover
in this scheme the limit $m_Q \to \infty$ can be performed
in a well defined way. On the other
hand it might be argued that in the decay
$B^{**} \to B \pi$ the pion momentum is not constrained to be small, since
$m_{B^{**}} - m_B \simeq \Delta \simeq 500$ MeV. We shall comment
on the uncertainties introduced by
a small pion momentum  approximation in the next Section,
where we shall present a light-cone sum rules
calculation of $G_{B^{**} B \pi}$ which is not based on S.P.L.

The problem arising in the soft pion limit is related to the fact that
$q=0$ implies $q_1^2 \, = \, q_2^2$. As a consequence, one cannot perform
a double Borel transform in the variables $q_1^2 \, , \, q_2^2$, and the
single Borelization procedure  has to be used.  As
well known \cite{eletski} in this way
one introduces unwanted not-exponentially suppressed contributions
(the so called {\it parasitic terms}) that have to be
estimated\footnote{A similar situation is met computing
$g_{B^*B\pi}$ by this method
\cite{g}.}. We shall show in the
sequel how this problem can be solved. For
the time being we consider the result of computing
$B(q_1^2, q_1^2, 0)$  by OPE in the soft pion limit.
The result of a straightforward analysis (much similar to that considered
in Refs.\cite{g,hyperfine}) is as follows:
\begin{equation}
B(q_1^2, q_1^2, 0) =B^{(0)}+B^{(1)}+B^{(2)}+B^{(3)}+B^{(4)}+B^{(5)}
\label{terms}
\end{equation}
where
\begin{eqnarray}
B^{(0)}&=&-\frac{<{\overline u}u>}{f_\pi} \frac{1}{q_1^2-m_b^2} \nonumber\\
B^{(1)}&=&0 \nonumber\\
B^{(2)}&=&-\frac{m_0^2 <{\overline u}u>}{4 f_{\pi}  (q_1^2-m_b^2)^2} \left [
1 - \frac{2 m_b^2}{q_1^2-m_b^2} \right] \nonumber\\
B^{(3)}&=&0 \nonumber\\
B^{(4)}&=&{{m_0^2 <{\overline u}u>} \over{4 f_{\pi}
(q_1^2-m_b^2)^2}} \nonumber\\
B^{(5)}&=&0 \; . \label{cntr}
\end{eqnarray}
In Eqs. (\ref{cntr}) $<{\overline u}u>$ is
the quark condensate ($<{\overline u}u> =-(240 MeV)^3$),
$m_0^2$
is defined by the equation
\begin{equation}
<{\overline u} g_s \sigma \cdot G u> =m_0^2<{\overline u}u> \;
\end{equation}
and its numerical values is: $m_0^2=0.8 \; GeV^2$.
The origin of the different terms in Eq.(\ref{terms}) is as follows.
$B^{(0)}$ is the leading term in the short distance expansion; $B^{(1)}$,
$B^{(2)}$ and $B^{(3)}$ arise from the expansion of $j_5(x)$ at the
first, second and third order in powers of $x$; $B^{(4)}$ and $B^{(5)}$ arise
 from the expansion of the heavy quark propagator at the second order and from
the zeroth and first term in the expansion of $j_5(x)$ respectively. We
have considered all the operators with dimension  $D\le 5$ in
the O.P.E. of the currents appearing in Eq.(\ref{corr1}).

Let us now compute the hadronic side
of the sum rule, that we call $B_{had}$.
We divide the integration region into two regions $D_1, D_2$,
as depicted in Fig. 1. $D_1$ is bounded by the lines $s=m_b^2, s'=m_b^2$ and
$s+s'=C$. We assume that $C$ satisfies the bounds:
\begin{equation}
min({s'}_0+m^2_{B^{**}},{s}_0+m^2_B)
\geq C \geq m^2_{B^{**}}+m^2_B \label{bound}
\end{equation}
where $s_0$ and ${s'}_0$ are thresholds for continuum production in the
variables $s$ and $s'$ respectively.
The  bound in (\ref{bound})  is chosen in such a way that, inside $D_1$,
only the poles $B^*$ and $B^{**}$ (in the variable $s$) and $B$
(in the variable $s'$) are included. Their contribution is as follows:
\begin{equation}
B_{pole}(q_1^2,q_1^2,0)=-
\frac{G_{B^{*} B \pi} f_{B} f_{B^*} m_B
(m^2_{B^*}-m^2_B)}{4 m_b m_{B^*}(q_1^2-m^2_{B^*})
(q_1^2-m^2_B)}  -
\frac{G_{B^{**} B \pi} f_{B} f_{B^{**}} m^2_B}{2 m_b (q_1^2-m^2_{B^{**}})
(q_1^2-m^2_B)} \; .\label{bp}
\end{equation}
We observe that, due to the presence of the factor
$m^2_{B^*}-m^2_B$ the contribution of the $1^-$ pole is strongly
depressed as compared to the term describing the $0^+$ resonance;
in the $m_b \to \infty$ limit, the ratio of the
two terms is $O(\frac{1}{m^2_b})$. We have checked that this suppression
holds not only for the beauty, but also for the charm, where the
vector meson pole contribution
is less than  $3\%$ of the $0^+$ pole. For this reason
we shall omit the $1^-$ pole in the sequel and we shall take only the
scalar contribution in Eq.(\ref{bp}).

Let us now consider the region $D_2$, where a continuum of resonances
contributes. Following Ref. \cite{neubprep}
we assume the following model for the hadronic
dispersive function in this region:
\begin{equation}
\rho (s,s') \; = \; f (s,s') \delta(s-s')
\theta(s \, + \, s'\, -\, C)~~~ ~~~~~~~~~(region \; \; D_2) \; .
\label {d2}
\end{equation}
In order to justify this choice let
us introduce the variables $z=s-s'$ and $y=\frac{s+s'}{2}$.
In these new variables we write the hadronic side
of the sum rule as follows:
\begin{equation}
B_{had}(q^2_1,q^2_1,0)=B_{pole}(q^2_1,q^2_1,0)+
B_{cont}(q_1^2)
\end{equation}
where the contribution of the continuum of resonances (region $D_2$)
is as follows:
\begin{equation}
B_{cont}(q_1^2)=\frac{1}{\pi^2}
\int_{C/2}^{\infty} dy
\int_{-2(y-m_b^2)}^{+2(y-m_b^2)} dz
\frac{1}{(y+\frac{z}{2}-q_1^2)(y-\frac{z}{2}-q_1^2)}
\rho (y+ \frac{z}{2},y - \frac{z}{2}) \; . \label{f}
\end{equation}
Let us now rewrite the
leading term in the O.P.E. expansion as an integral over the variables
$z, \; y$:
\begin{eqnarray}
B^{(0)} &=&- {{ <{\overline u}u>}\over{f_{\pi}}}
{{1}\over{q_1^2-m_b^2}}  \; = \nonumber\\
&=& {{ <{\overline u}u>}\over{f_{\pi}}}
\int_{m_b^2}^{\infty} dy {{1}\over{(y \, - \, q_1^2)^2}}
\int_{-2(y-m_b^2)}^{+2(y-m_b^2)} dz \delta (z) \; . \label{b2}
\end{eqnarray}
To relate (\ref{f}) and (\ref{b2}), one
imposes local duality in $y$  \cite{neubprep}; this means
that inside the region $D_2$, i.e. for $y \geq C/2$, the arguments of the
$y$-integrals in Eqs. (\ref{f}) and (\ref{b2}) should be equal. This
implies that (\ref{d2}) is valid with
\begin{equation}
f (s,s') \; = {\pi^2} {{ <{\overline u}u>}\over{f_{\pi}}}
\end{equation}
which fixes the model of the hadronic continuum in the region $D_2$.
A justification for local duality in the variable $y$ can be
obtained by explicit calculations in particular models \cite{shifma}.

Using these results we obtain, after the Borel transform, the following sum
rule, which as we have already observed, is valid only in the soft pion limit:
\begin{eqnarray}
\frac{G_{B^{**} B \pi} f_{B} f_{B^{**}} m^2_B }{m^2_{B^{**}}-m^2_B} &&
\left [  e^{-m_B^2/M^2} -
 e^{-m_{B^{**}}^2/M^2} \right]
\nonumber\\
&=& -{{2 m_b <{\overline u}u>}\over{f_{\pi}}} \left [
e^{-m_b^2/M^2} \left [ 1- \frac{m_b^2 m_0^2}{4 M^4}
\right] - e^{C/(2 M^2)}\right] \; . \nonumber\\
\label{sr}
\end{eqnarray}
The sum rule is valid provided one imposes two conditions on the Borel
parameter $M^2$. First one requires that the contribution of the
continuum does not exceede the pole contributions (for larger values
this would produce uncontrollable uncertainties in $G_{B^{**} B \pi}$ due to
our poor knowledge of the continuum); this fixes the upper bound for
$M^2$. On the other hand, for the O.P.E to be meaningful,
we have to impose that the higher power terms in $1/M^2$ in
Eq.(\ref{sr}) have decreasing values, which fixes the lower bound for $M^2$.
 For the beauty sector we use the values
$m_{B^{**}}-m_B = \Delta \approx 500 \pm 100$ MeV, $m_b= 4.6\; GeV$, the
parameter C in the range $(61 - 64) \; GeV^2$ (with thresholds
$s_0={s'}_0 \simeq 36 \; GeV^2$). The previous
criteria are satisfied for $M^2$ in the range
$(15-25) \; GeV^2$. We get the result:
\begin{equation}
G_{B^{**} B \pi} f_{B} f_{B^{**}} \; = \; 0.43 \pm 0.06 \; GeV^3 \; .
\end{equation}

For  charm, we
use
$m_{D^{**}}- m_D \approx \Delta =500 \pm 100$ MeV, $m_c= 1.35 \; GeV$, and the
parameter C in the range$(10 - 14) \; GeV^2$ (with thresholds
$s_0={s'}_0 \simeq 6 \; GeV^2$). The
criteria for $M^2$ are satisfied in the range
$M^2 = (2-8)$ GeV$^2$, and one gets the result:
\begin{equation}
G_{D^{**} D \pi} f_{D} f_{D^{**}} = 0.38 \pm 0.11 \; GeV^3 \; .
\end{equation}

To get $G_{B^{**} B \pi} $ and $G_{D^{**} D \pi} $ we use
$f_B=0.18 \pm 0.03 \; GeV$ and
$f_{B^{**}}=0.18 \pm 0.03 \; GeV$ \cite{noi} obtaining:
\begin{equation}
G_{B^{**}B\pi}= 13.3 \pm 4.8 \; GeV \;,
\end{equation}
while, using
$f_D=0.195 \pm 0.020 \; GeV$ and
$f_{D^{**}}=0.17 \pm 0.02 \; GeV$, \cite{noi}  we obtain:
\begin{equation}
G_{D^{**}D\pi} = 11.5 \pm  4.0 \; GeV \; .
\end{equation}

We observe substantial violations of the scaling law
$G_{D^{**}D\pi}/G_{B^{**}B\pi} \approx \frac{m_c}{m_b}$. On the
other hand the ratio $R=\frac{G_{B^{**} B \pi} f_{B} f_{B^{**}}}{G_{D^{**}
D \pi} f_{D}
 f_{D^{**}}}$ is less sensitive
to scaling violations (numerically we find $R \simeq 1.13$, to
be compared to the scaling prediction $R=1$).

Let us now take the limit $m_b \to \infty$. This limit is defined
by Eqs.
(\ref{para})  and by a rescaling of the Borel parameter $M^2$:
\begin{equation}
M^2\; = \; 2 m_b \, E \; .
\end{equation}
In this way we obtain the asymptotic rule:
\begin{equation}
h \hat F {\hat F}^+ \left [ 1 \; - \; e^{-\Delta/E} \right] =
4 <{\overline u}u> \left [
e^{\omega/E} (1\; - \; \frac{m_0^2}{16 E^2}) \; - \; e^{-\delta/E} \right] \; ;
\end{equation}
the quantities $\delta, \; {\hat F}$ and ${\hat F}^+$, defined by:
\begin{eqnarray}
\delta &=&\frac{1}{2 m_b} \left [ \frac{C}{2} - m_B^2 \right] \nonumber\\
{\hat F} &=& f_B \sqrt{ m_B} \nonumber \\
{\hat F}^+  &=& f_{B^{**}} \sqrt{ m_{B^{**}}} \label{coup}
\end{eqnarray}
remain finite in the infinite heavy quark mass limit, modulo
logarithmic corrections.
The constraints on the Borel parameter imply that
$E$ must be in the range $ (0.5 - 1.4) \; GeV$.
Using $\omega= 0.50 \pm 0.07$ GeV,
$\Delta= 500$ MeV, and $\delta = 400 \pm 50$ MeV
the numerical outcome of the sum rule is:
\begin{equation}
h \hat F {\hat F}^+ \; = \, - \; 0.072 \pm 0.008 \; GeV^3 \label{fres}
\end{equation}
(the results are weakly dependent on $\delta$).
By the values $\hat F = 0.30 \pm 0.05$ GeV$^{3/2}$ \cite{neub}
and ${\hat F}^+ \; = \; 0.46\pm 0.06$ GeV$^{3/2}$ \cite{noi}  we obtain:
\begin{equation}
h=\, - \, 0.52 \pm 0.17 \; . \label{h1}
\end{equation}
\section{A light-cone sum rules calculation of \lowercase{$h$}}

An independent calculation of the strong coupling $h$ can be carried out
using a method that allows us to
consider arbitrary momenta of the pion.
We shall consider the correlation function
\begin{equation}
A(q_1^2, q_2^2, q^2) = i \int dx <\pi^+(q)| T( j_5(x) j(0) |0> e^{-i q_2 x}
\label{corr2}
\end{equation}
where $j_5=i{\overline u} \gamma_5 d$ (as before) and $j ={\overline b} d$.

The method consists in expanding the T-product of the quark currents,
appearing in Eq.(\ref{corr2}), near the light-cone, in terms of
non-local operators whose matrix elements can be written as wave functions of
increasing twist.

This approach finds its origin in the analysis
of hard exclusive processes in
QCD \cite{chern1}; within this framework, strong
couplings ($g_{\omega \rho \pi}$, $g_{\pi N N}$), form factors (such as $\pi A
\gamma^*$, the pion form factor at intermediate momentum transferred)
and nucleon magnetic moments
have been calculated \cite{chern2}.
Light-cone sum rules have also been used to calculate the
form factors governing
$B$ and $D$ meson semileptonic decays \cite{ball},  and the radiative
$B \to K^* \gamma$ transition \cite{ali}.

In our calculation of $h$ we
follow the notations adopted by the recent paper by
Belyaev et al. \cite{ruckl} devoted to the calculation
of the coupling constants $g_{B^* B \pi}$ and $g_{D^* D \pi}$.
Within the light-cone sum rules approach
the correlator Eq.(\ref{corr2}) can be written as follows:
\begin{eqnarray}
& &A^{QCD}(q^2_1,q^2_2,q^2)
=-{2 m_b <{\bar u} u> \over f_{\pi}} \int_0^1 du
{\varphi_P(u) \over m_b^2 -(q_2+uq)^2} +  \nonumber \\
&+&f_{\pi}(q_2 \cdot q) \int_0^1 du
\Big\{ {\varphi_{\pi}(u) \over m_b^2 -(q_2+uq)^2}
-{4g_1(u) \over [m_b^2 -(q_2+uq)^2]^2}
\Big(1+ {2 m_b^2 \over m_b^2 -(q_2+uq)^2} \Big) \Big\} \nonumber \\
&+& f_{\pi}(q_2 \cdot q) \int_0^1 du
\int {\cal{D}}\alpha_i {(1-2u)\chi(\alpha_i)+\tilde{\chi}(\alpha_i) \over
[m_b^2 -(q_2+q(\alpha_1+u \alpha_3))^2]^2} \hskip 3 pt .\label{lc}
\end{eqnarray}

The pion wave functions $\varphi_P(u)$, $\varphi_\pi(u)$ and $g_1(u)$
appear in the matrix elements of non-local quark operators
\cite{ruckl}:
\begin{eqnarray}
<\pi(q)| {\bar d} (x) \gamma_{\mu} \gamma_5 u(0) |0>&=&-i f_{\pi} q_{\mu}
\int_0^1 du e^{iuqx} (\varphi_{\pi}(u) +x^2 g_1(u) + {\cal O}(x^4) )
\nonumber \\
&+& f_\pi \big( x_\mu - {x^2 q_\mu \over q x} \big)
\int_0^1 du e^{iuqx}  g_2(u) \hskip 3 pt  , \label{ax} \\
<\pi(q)| {\bar d} (x) i \gamma_5 u(0) |0> &=& {f_{\pi} m_{\pi}^2 \over m_u+m_d}
\int_0^1 du e^{iuqx} \varphi_P(u)  \hskip 3 pt ,
 \label{pscal}
\end{eqnarray}
where ${f_{\pi} m_{\pi}^2 \over m_u+m_d}$ is related to
the light-quark condensate $<\bar u u>$ by the current algebra
relation ${f_{\pi} m_{\pi}^2 \over m_u+m_d}=-2 {<\bar u u>\over f_\pi}$.
It should be noticed that the path-ordered gauge factor
$P \exp\big(i g_s \int_0^1 du x^\mu A_\mu(u x) \big)$ has been omitted
in the above matrix elements due to the choice of the
light-cone gauge  $x^\mu A_\mu(x) =0$.

The functions
$\chi(\alpha_1, \alpha_2, \alpha_3)$ and
$\tilde{\chi}(\alpha_1, \alpha_2, \alpha_3)$ are combinations of
twist-four wave functions
$\varphi(\alpha_1, \alpha_2, \alpha_3)$ and
$\tilde \varphi(\alpha_1, \alpha_2, \alpha_3)$ \cite{fil}:
\begin{equation}
\chi(\alpha_1, \alpha_2, \alpha_3)=2 \varphi_{\bot}
(\alpha_1,\alpha_2,\alpha_3)-
 \varphi_{\|}(\alpha_1,\alpha_2,\alpha_3) \hskip 3 pt \label{chi}
\end{equation}
and
\begin{equation}
\tilde{\chi}(\alpha_1, \alpha_2, \alpha_3)=2 \tilde\varphi_{\bot}
(\alpha_1,\alpha_2,\alpha_3)-
\tilde\varphi_{\|}(\alpha_1,\alpha_2,\alpha_3) \hskip 3 pt,
\label{chitilde}
\end{equation}
\noindent
where the functions $\varphi$ and $\tilde \varphi$ parametrize
the matrix elements of quark-gluon
operators:
\begin{eqnarray}
& &<\pi(q)| {\bar d} (x)
\gamma_{\mu} \gamma_5 g_s G_{\alpha \beta}(ux)u(0) |0>=
\nonumber \\
&&f_{\pi} \Big[ q_{\beta} \Big( g_{\alpha \mu}-{x_{\alpha}q_{\mu} \over q \cdot
x} \Big) -q_{\alpha} \Big( g_{\beta \mu}-{x_{\beta}q_{\mu} \over q \cdot x}
\Big) \Big] \int {\cal{D}} \alpha_i \varphi_{\bot}(\alpha_i)
e^{iqx(\alpha_1 +u \alpha_3)}\nonumber \\
&&+f_{\pi} {q_{\mu} \over q \cdot x } (q_{\alpha} x_{\beta}-q_{\beta}
x_{\alpha}) \int {\cal{D}} \alpha_i \varphi_{\|} (\alpha_i)
e^{iqx(\alpha_1 +u \alpha_3)} \hskip 3 pt  \label{gi}
\end{eqnarray}
\noindent and
\begin{eqnarray}
& &<\pi(q)| {\bar d} (x) \gamma_{\mu}  g_s \tilde G_{\alpha \beta}(ux)u(0) |0>=
\nonumber \\
&&i f_{\pi}
\Big[ q_{\beta} \Big( g_{\alpha \mu}-{x_{\alpha}q_{\mu} \over q \cdot
x} \Big) -q_{\alpha} \Big( g_{\beta \mu}-{x_{\beta}q_{\mu} \over q \cdot x}
\Big) \Big] \int {\cal{D}} \alpha_i \tilde \varphi_{\bot}(\alpha_i)
e^{iqx(\alpha_1 +u \alpha_3)}\nonumber \\
&&+i f_{\pi} {q_{\mu} \over q \cdot x } (q_{\alpha} x_{\beta}-q_{\beta}
x_{\alpha}) \int {\cal{D}} \alpha_i \tilde \varphi_{\|} (\alpha_i)
e^{iqx(\alpha_1 +u \alpha_3)} \hskip 3 pt . \label{git}
\end{eqnarray}
\noindent In Eq. (\ref{git}) the field
$\tilde G_{\alpha \beta}$  is the dual of $G_{\alpha \beta}$:
$\tilde G_{\alpha \beta}= {1\over 2} \epsilon_{\alpha \beta \delta \rho}
G^{\delta \rho} $; the integration on the variables $\alpha_i$ is performed
considering that
${\cal{D}} \alpha_i =d \alpha_1 d \alpha_2 d \alpha_3
\delta(1-\alpha_1 -\alpha_2
-\alpha_3)$.

Using the normalization of the function $\varphi_P$:
$\int_0^1 \varphi_P(u) du=1$
we recover from Eq. (\ref{lc}),
in the soft pion limit $q \to 0$,
 the leading order expansion of
$B(q^2_1,q^2_1,0)$ in Eq. (\ref{cntr}), apart from
an overall factor $- 2 m_b$ due to the the different choice
of the current interpolating the $B^{**}$ in the two cases (vector and
scalar, respectively); it should be noticed that
the terms proportional to the mixed
quark-gluon condensate appearing in Eq. (\ref{cntr})
are missing here since they are related to
higher-twist  pion wave functions.

The hadronic representation of
$A(q^2_1,q^2_2,q^2)$ Eq. (\ref{corr2})
can be expressed in terms of the contribution of the lowest-lying resonances
$B$ and $B^{**}$:
\begin{equation}
A_{pole}(q^2_1,q^2_2,q^2)=
G_{B^{**} B \pi} f_B f_{B^{**}} {m^2_B m^2_{B^{**}} \over
(m_b+m_d) (m_b-m_u)}
{1 \over (m^2_{B^{**}}- q_1^2)(m^2_B- q_2^2)}
\end{equation}
in the region $m_b^2 \le s \le s_0$,
$m_b^2 \le s' \le s'_0$,
and of the contribution of higher states and of the hadronic continuum.
For $q_1 \ne q_2$ we can perform a double independent borelization
in the variables
$-q_1^2$ and $-q_2^2$. In this way, the parasitic
contributions coming from the resonance-continuum terms in (\ref{corr2})
are exponentially suppressed. From Eq. (\ref{corr2}) we get:
\begin{equation}
{\cal{B}} \; A_{had} = {1 \over M_1^2 M_2^2} {1 \over \pi^2}
\int ds \; ds' \rho_{had}(s, s', q^2) \exp( - {s\over M_1^2}- {s'\over M_2^2})
\end{equation}
where $M_1^2$ is the Borel parameter associated to the variable
$-q_1^2$ and $M_2^2$ to $-q_2^2$. In this expression the
contribution of the pole reads:
\begin{equation}
{\cal{B}} \; A_{pole} = {1 \over M_1^2 M_2^2}
G_{B^{**} B \pi} f_B f_{B^{**}} {m^2_B m^2_{B^{**}} \over
(m_b+m_d) (m_b-m_u)}
   \exp \Big[-{m_{B^{**}}^2\over M_1^2} -{m_B^2\over M_2^2} \Big] \; .
\end{equation}
On the other hand, borelization of
$A^{QCD}$ provides us with the following expression:
\begin{eqnarray}
 {\cal{B}} \; A^{QCD}
&=&{1 \over M_1^2 M_2^2} \exp \Big[ -{m_b^2 \over M^2} \Big]
\Big\{-2{m_b <{\bar u}u> \over f_{\pi}}  M^2 \varphi_P(u_0)\nonumber \\
&-&f_{\pi} {M^4 \over 2} \varphi_{\pi}^{\prime}(u_0)
+2 f_{\pi}(M^2+m_b^2)
g_1^{\prime}(u_0) \nonumber \\
&+&f_{\pi} {M^2 \over 2} \Big[ \int_0^{u_0} {d \alpha_3 \over \alpha_3}
[\tilde \chi(u_0-\alpha_3, 1-u_0, \alpha_3) -
\chi(u_0-\alpha_3, 1-u_0, \alpha_3)] \nonumber \\
&-&\int_0^1 {d \alpha_3 \over \alpha_3}
(\chi(u_0, 1-u_0-\alpha_3, \alpha_3)
+ \tilde\chi(u_0, 1-u_0-\alpha_3, \alpha_3) ) \nonumber \\
&+&2\int_0^{u_0}d \alpha_1 \int_{u_0-\alpha_1}^{1-\alpha_1}
{d \alpha_3 \over \alpha_3^2}
{\chi}(\alpha_1,1-\alpha_1-\alpha_3,\alpha_3) \Big] \Big\} \; ,
\label{borel}
\end{eqnarray}
where
$M^2= {M_1^2 M_2^2 \over M_2^2 +M_2^2}$ and
$u_0= {M_1^2 \over M_2^2 +M_2^2}$,
$\varphi'= {d \varphi \over du}$ and
$g_1'= {d g_1 \over du}$. It is worth observing that, choosing
the symmetric point $u_0={1 \over 2}$ (which corresponds to
a quark and an antiquark of the same momentum inside the pion)
Eq. (\ref{borel}) is considerably simple, since at this point both
$\varphi'$ and $g_1'$ vanish.
Moreover, at $u_0={1 \over 2}$ the subtraction of the
continuum contribution can be done by substituting
$e^{-{m_b^2 \over M^2}} \to
 e^{-{m_b^2 \over M^2}} - e^{-{s_0 \over M^2}}$,
at least in the twist 3 contribution
\cite{ruckl} (we use this substitution everywhere in
Eq. (\ref{borel}) since higher twist contributions are numerically small).
Therefore, we derive the sum rule:
\begin{eqnarray}
&& G_{B^{**}B\pi} f_B f_{B^{**}} {m_B^2 m_{B^{**}}^2 \over (m_b+m_d) (m_b-m_u)}
   \exp \Big[-{m_{B^{**}}^2+m_B^2 \over 2 M^2} \Big]= \nonumber \\
&=&\Big\{\exp \Big[-{m_b^2 \over M^2} \Big]-\exp \Big[-{s_0 \over M^2} \Big]
\Big\} {M^2 f_{\pi} \over 2}
 \Big\{ -{4 m_b <{\bar u } u > \over f_{\pi}^2} \varphi_P(u_0)
\nonumber \\
&+& \int_0^{u_0} {d \alpha_3 \over \alpha_3}
[\tilde\chi(u_0-\alpha_3, 1-u_0,\alpha_3) -
 \chi(u_0-\alpha_3, 1-u_0,\alpha_3)] \nonumber \\
&-& \int_0^1 {d \alpha_3 \over \alpha_3}
(\chi(u_0, 1-u_0-\alpha_3,\alpha_3)) +
\tilde \chi(u_0, 1-u_0-\alpha_3,\alpha_3)) )\nonumber \\
&+& 2
\int_0^{u_0}d \alpha_1 \int_{u_0-\alpha_1}^{1-\alpha_1}
{d \alpha_3 \over \alpha_3^2}
\chi(\alpha_1, 1-\alpha_1-\alpha_3,\alpha_3) \Big\} \; .
\label{sumrule}
\end{eqnarray}

Let us now discuss the numerical analysis of Eq.(\ref{sumrule}).
In the $b$-channel we choose the same values for $m_b$,
$m_{B^{**}}$ and $<\bar u u>$ used in Section III.
The main nonperturbative quantities are
the twist 3 function $\varphi_P$ and the combinations $\chi$ and $\tilde \chi$
of twist 4 wave-functions. We choose the model in Ref. \cite{fil}, where such
quantities have been fixed in the framework of a systematic expansion in the
conformal spin. It turns out that, using
$\varphi_P(1/2)|_{\mu_b}=1.07$ \cite{fil}, the higher twist contribution in
Eq.(\ref{sumrule}) is $2 \%$ of the twist 3 contribution,
and therefore the numerical value of $\varphi_P(1/2)$ represents
a crucial quantity in our analysis. We allow the effective threshold
$s_0$ to vary in the range $36 - 40 \; GeV^2$. Moreover, we fix
the highest value of the Borel parameter $M^2$ in the duality window
by imposing that the contribution
of the continuum is $30 \%$ of the resonance. In this way we find
$M^2_{max}=12-14 \; GeV^2$. The minimum value of $M^2$
is usually fixed by imposing that terms proportional
to higher powers of $1/M^2$ are small enough. Since in Eq.(\ref{sumrule})
such terms are absent, we only look for a stability region in $M^2$, and
choose $M^2_{min}=6 \; GeV^2$, which is the same value
adopted in the analysis
of $g_{B^*B\pi}$. With these input parameters we find:
\begin{equation}
G_{B^{**}B\pi} f_B f_{B^{**}} = 0.69 \pm 0.14 \; GeV^3
\end{equation}
where the uncertainty is due to the variation of $s_0$ and
to the dependence of the numerical results on $M^2$. Therefore, using
the same values of the leptonic constants
$f_B$ and $f_{B^{**}}$
adopted in the previous Section,
we get:
$G_{B^{**}B\pi}= 21 \pm 7 \; GeV$ and, from (\ref{grel}),
$h(m_b)=- \, 0.52 \pm 0.18 $.

In the case of the charm sector, we use
$\varphi_P(1/2)|_{\mu_c}=1.14$, $s_0=9 - 11 \; GeV^2$ and $M^2$ in the range
$M^2=2-5 \; GeV^2$. We obtain:
\begin{equation}
G_{D^{**}D\pi} f_D f_{D^{**}} = 0.21 \pm 0.02 \; GeV^3 \;
\end{equation}
and the results:
$G_{D^{**}D\pi} = 6.3 \pm 1.2 \; GeV$
(using $f_D$ and $f_{D^{**}}$ as in Section III) and
$h(m_c)=-0.44 \pm 0.09 $.

A two-parameter fit of the above results in the form
\begin{equation} h(m)=h \; (1 + {\sigma \over m}) \label{fit}
\end{equation}
gives the result:
\begin{equation}
h=-0.56 \pm 0.28 \label{h2}
\end{equation}
 and for the parameter $\sigma$,
$\sigma=0.4 \pm 0.8 \; GeV$.

Let us now compare these results with those of the previous Section. The values
of $h$ found by the two methods agree with each other.
As for the finite mass results, the two methods
sensibly differ (almost a factor of 2) in the case
of the charm, while the deviation is
less important for the case of beauty (around $40\%$). These
differences should be attributed to
corrections to the soft pion limit that have not been
incorporated in the
results of Section III.
\section{Excited heavy mesons widths}

In this Section we apply our results for the strong coupling constants
to the calculation of the hadronic widths of the excited mesons.
First of all, we can compute the strong widths of the decays
$P_0 \to P \pi$ and $P_1 \to P^{*} \pi$. Differently from the decays
of the positive parity states having $s_{\ell}^+={\frac{3}{2}}^+$, i.e. the
states $P_2$ and $P_{1}'$, where the final pion is in $D$-wave, these are
$S$-wave decays; moreover,
as shown in \cite{falkluke},
the single pion channels are expected to saturate the total widths.

In the $m_Q \to \infty$ limit one would obtain
\begin{equation}
\Gamma( P_0 \to P^+ \pi^-)= \Gamma( P_1 \to P^{*+} \pi^-)=
\frac{1}{2\pi} \left( \frac{h}{f_\pi } \right)^2 \Delta^3
\label{1}
\end{equation}
but this formula is of limited significance,
especially for the case of charm, due to the large $1/m_Q$ corrections
coming from the kinematical factors.

Keeping $m_Q$ finite, the formulae become:
\begin{equation}
\Gamma (P_0 \to P^+ \pi^-) =
\frac{1}{8\pi} G_{P^{**}P\pi}^2 \frac{\left[ (m^2_{P_0} -
(m_P + m_{\pi})^2) (m^2_{P_0} -
(m_P - m_{\pi})^2)\right]^{\frac{1}{2}}}{2 m_{P_0}^3} \; .
\label{3}
\end{equation}
Using $G_{D^*D\pi}=6.3 \pm 1.2 \; GeV$,
$G_{B^*B\pi}=21\pm 7 \; GeV$, and
$\Delta_D = \Delta_B = 500 \; MeV$, one finds
\begin{eqnarray}
\Gamma(D_0 \to D \pi) & \simeq & 180 \; MeV \\
\Gamma(B_0 \to B \pi) & \simeq & 360 \; MeV \; .
\label{4}
\end{eqnarray}

For the decay $1^+ \to 1^- \pi$ ($P_1 \to P^* \pi$) we find
\begin{eqnarray}
\Gamma (P_1 \to P^{*+} \pi^-) & = &\frac{G_{P_1P^*\pi}^2}{8\pi}
\frac{\left[ (m^2_{P_1} -
(m_{P^*} + m_{\pi})^2) (m^2_{P_1} -
(m_{P^*} - m_{\pi})^2)\right]^{\frac{1}{2}}}{2 m_{P_1}^3} \times \nonumber \\
& \times & \frac{1}{3} \left( 2+ \frac{(m^2_{P_1}
+ m^2_{P^*})^2}{4 m^2_{P_1} m^2_{P^*}}
\right) \; .
\label{5}
\end{eqnarray}
In the limit $m_Q \to \infty$ Eqs.(\ref{5}) and (\ref{3}) coincide.
Notice that we have not computed the coupling $G_{P_1 P^* \pi}$.
In the infinite-mass limit it coincides with
$G_{P^{**}P \pi}$ and therefore, in order to
estimate the widths of the $1^+$ states, we assume that this equality
holds for finite mass as well. From Eq.(\ref{5}) we obtain:
\begin{eqnarray}
\Gamma(D_1 \to D^* \pi) & \simeq & 165 \; MeV \\
\Gamma(B_1 \to B^* \pi) & \simeq & 360 \; MeV \; .
\label{6}
\end{eqnarray}
Also in this case we have taken $m_{P_1} - m_{P^*} = 500 \;$ MeV ($P = B,D$)
as suggested by HQET considerations.

In order to perform some comparison with the experimental data, we
write down also the formulae giving the widths of the
$s_{\ell}^+={\frac{3}{2}}^+$ states \cite{falkluke}:
\begin{eqnarray}
\Gamma (P_2^0 \to P^+ \pi^-) & = & \frac{1}{15\pi} \frac{m_P}{m_{P_2}}
\frac{h'^2}{{\Lambda_\chi}^2} \frac{|\vec{p}_{\pi}|^5}{{f_\pi}^2} \\
\Gamma (P_2^0 \to P^{*+} \pi^-) & = & \frac{1}{10\pi} \frac{m_{P^*}}{m_{P_2}}
\frac{h'^2}{{\Lambda_\chi}^2}\frac{|\vec{p}_{\pi}|^5}{{f_\pi}^2} \\
\Gamma (P_1^{\prime 0} \to P^{*+} \pi^-) & = & \frac{1}{6\pi}
\frac{m_{P^*}}{m_{P_2}}
\frac{h'^2}{{\Lambda_\chi}^2}\frac{|\vec{p}_{\pi}|^5}{{f_\pi}^2} \; .
\label{tre}
\end{eqnarray}

The strong coupling constant $\frac{h'}{\Lambda_\chi}$ can be estimated
from the decay widths of the charmed state $D_2(2460) \to D \pi, D^* \pi$;
using $\Gamma_{tot}(D_2)=21 \pm 5$ MeV \cite{pdg94},
and assuming that only two body
decays are relevant, one gets $\frac{h'}{\Lambda_\chi} \approx
0.55$ GeV$^{-1}$. From this
result and from Eq. (\ref{tre}) one obtains for the state
$D_1^{\prime0}$
the total width $\Gamma_{tot}\approx 6$ MeV; on the
other hand the experimental width of the other narrow state observed
in the charm sector, i.e. the $1^+$ $D_1(2420) $ particle, is
$\Gamma_{tot} (D_1(2420)) = 18
\pm 5$ MeV \cite{pdg94}.
This discrepancy could be attributed to a mixing between
the $D_1$ and the $D_1'$ states \cite{iswi}.
If $\alpha$ is the mixing angle,
we have
\begin{equation}
\sin^2 (\alpha) \approx \frac{12 \; MeV}{\Gamma(D_1) - \Gamma(D_1')} \simeq
0.08
\label{7}
\end{equation}
and therefore we get the estimate $\alpha \approx 16^o$. This
determination agrees with the result of Kilian et al. in Ref.\cite{falkluke}.

As for the $B$ sector, evidence has been recently reported of a bunch of
positive parity states
$B^{**}$, with an average mass $m_{B^{**}} = 5732 \pm 5 \pm 20$ MeV
and an average width $\Gamma(B^{**}) = 145 \pm 28$ MeV \cite{delphi,opal}.
The observed states can be identified
with the two doublets $(2^+, 1^+)$ and $(1^+, 0^+)$. We note that
the mass splitting, $\Delta = 500 \; MeV$, between
$S$ and $P$ states that we have chosen,
agrees rather well with the experimental result in the
$B$ sector; our predictions for the widths, using the experimental value for
the $B^{**}$ mass,
are: $\Gamma_{tot}(B_0) \simeq 330 \; MeV$,
$\Gamma_{tot}(B_1) \simeq  300\; MeV$,
$\Gamma_{tot}(B_2) \simeq 12$ MeV and   $\Gamma_{tot}(B_1') \simeq 10$ MeV
(we have neglected here the mixing which
is a $1/m_Q$ effect).

It is difficult to perform a detailed comparison of
these results with the yet uncomplete experimental outcome;
however, assuming that
the result obtained by LEP collaborations represents an
average of several states, its value is compatible with our estimate
of the widths. Opal
 \cite{opal} has also reported evidence of a
$B^{**}_s$ state
with mass $m_{B^{**}_s} = 5853 \pm 15$ MeV  and width
$\Gamma_{B^{**}_s} = 47 \pm 22$ MeV. The width can be interpreted
as connected to the decay $B^{**}_s \to B K, \; B^* K$.
Assuming again that
the width is saturated by two-particle final states, and
using $m_{B^{**}_s} = 5853 \; MeV$,
we obtain:
\begin{eqnarray}
\Gamma(B^{**}_s(0^+)) & \simeq & 280 \; MeV \\
\Gamma(B^{**}_s \to B^* K) & \simeq & 200 \; MeV \; \; \; \; \; (s_{\ell}=
{\frac{1}{2}}^+) \\
\Gamma(B^{**}_s(1^+))  & \simeq & 0.45 \; MeV \; \; \; \; \; (s_{\ell}=
{\frac{3}{2}}^+) \\
\Gamma(B^{**}_s(2^+)) & \simeq & 1.4 \; MeV \; .
\end{eqnarray}
Also in this case a detailed comparison with the experimental results
cannot be performed without more precise measurements;
we  observe, however, that the
computed widths of the different
$B^{**}_s$ states are generally smaller than the corresponding
quantities of the $B^{**}$ particles, a feature which is reproduced by the
experiment.

\section{Excited states contribution to
$\lowercase{f}_{D_s}/\lowercase{f}_D$}

Aim of this section is to study the contribution of the excited heavy mesons
to the ratio of leptonic decay constants $f_{D_s}/f_D$.
They are defined as
\begin{equation}
<0| {\bar d} \gamma_\mu \gamma_5 c |D^+ (p)> = i f_D p_\mu
\end{equation}
\begin{equation}
<0| {\bar s} \gamma_\mu \gamma_5 c |D_s (p)> = i f_{D_s} p_\mu
\end{equation}

In the chiral $SU(3)$ limit such a ratio is one, and the chiral corrections
are expected to be of the order $m_s/\Lambda_{\chi}$.
In terms of the fields $H_a$ and $\xi$ defined in Section II, the left-handed
current ${\bar q}_a \gamma_\mu (1 - \gamma_5 ) Q$ corresponds to \cite{chirlag}
\begin{equation}
L^a_\mu = \frac{i}{2} {\hat F} < \gamma_\mu (1- \gamma_5 ) H_b
\xi^{\dagger}_{ba} > + \ldots \label{wcurr}
\end{equation}
where the dots denote terms of higher order in the heavy mass and chiral
expansion. At the lowest order, $f_{D_s} = f_{D} = {\hat F}/\sqrt{m_D}$.
The effective couplings of the higher order terms, contributing to
$SU(3)$-violating corrections to the ratio $f_{D_s}/f_D$, are unknown.
In addition, there are non-analytic corrections arising from chiral loops:
in previous works \cite{grin,goity} the one-loop ``log-enhanced''
terms of the form $m^2 \log (m^2/\mu^2)$ ($m = m_\pi$, $m_K$ or $m_\eta$) were
kept, giving
\begin{equation}
\frac{f_{D_s}}{f_D}= 1 + 0.07 + 0.21 g^2
\label{gg}
\end{equation}
The  effective coupling $g$, appearing in the heavy-light chiral lagrangian
(\ref{L}), gives the vertex $D^* D \pi$, appearing in the loops.

The loop corrections depend on an arbitrary renormalization point $\mu$: this
dependence is canceled by the $\mu$-dependence of the coefficients of higher
order operators, which are here neglected. When $\mu$ is of the order of the
chiral symmetry breaking scale $\Lambda_\chi \approx 1 \; GeV$, these
higher terms do not contain large logarithms and are supposed to be small
compared with the ones coming from the chiral loops.

The excited positive parity heavy mesons  contribute
to $SU(3)$ violating effects as virtual
intermediate states in chiral loops.
In Ref.\cite{falk} the ``log-enhanced'' terms due to these excited-states loops
has been computed: as we will see below, some of them are proportional
to $h^2$ and others depend linearly on $h$. It has been  pointed out that these
terms could be numerically relevant and could invalidate the chiral
estimate based only on the states $D$ and $D^*$; as we shall see below,
however, the
terms ${\cal O}(h^2 )$ and ${\cal O}( h )$, while important, tend to cancel.

In the following, we will present an independent calculation of the chiral loop
contributions to the ratio of leptonic decay constants, and we will give a
numerical estimate obtained by using
the QCD sum rules results for the couplings $g$ and $h$.

The chiral loop induced corrections to $f_D$ and $f_{D_s}$ come from the
diagrams of figs. 2,3 and 4.

The self-energy diagrams (fig. 2) give the following
wave function renormalization
factors:
\begin{eqnarray}
Z_D & =& 1 - \frac{3 g_D^2}{16 \pi^2 f_\pi^2} \left[3/2
C_1(\Delta_{D^*D},\Delta_{D^*D},m_\pi ) +
C_1(\Delta_{D^*_s D},\Delta_{D^*_s D},m_K ) +
\frac{1}{6} C_1(\Delta_{D^*D},\Delta_{D^*D},m_\eta ) \right] \nonumber \\
& + &  \frac{ h_D^2}{16 \pi^2 f_\pi^2} \left[3/2
C(\Delta_{P_0 P},\Delta_{P_0 P},m_\pi ) +
C(\Delta_{P_{0s} P},\Delta_{P_{0s} P},m_K ) + \right. \nonumber \\
& + & \left. \frac{1}{6}
C(\Delta_{P_0 P},\Delta_{P_0 P},m_\eta ) \right]
\end{eqnarray}
\begin{eqnarray}
Z_{D_s} & = & 1 - \frac{3 g_D^2}{16 \pi^2 f_\pi^2} \left[
2 C_1(\Delta_{D^* D_s},\Delta_{D^* D_s},m_K )
+ \frac{2}{3} C_1(\Delta_{D^*D},\Delta_{D^*D},m_\eta ) \right] \nonumber \\
& + &  \frac{ h_D^2}{16 \pi^2 f_\pi^2} \left[
2 C(\Delta_{P_{0} P_s},\Delta_{P_{0} P_s},m_K ) + \frac{2}{3}
C(\Delta_{P_0 P},\Delta_{P_0 P},m_\eta ) \right]
\end{eqnarray}
where the mass splittings $\Delta_{P^* P}= M_{P^*}- M_{P}$,
$\Delta_{P^* P_s}= M_{P^*}- M_{P_s}$ and
$\Delta_{P_s^* P}= M_{P_s^*}- M_{P}$  are  ${\cal O}(1/m_Q)$, while the mass
splittings
$\Delta_{P_0 P}= M_{P_0}- M_{P}$,
$\Delta_{P_{0s} P}= M_{P_{0s}}- M_{P}$ and
$\Delta_{P_0 P_s}= M_{P_0}- M_{P_s}$ between excited and ground states
are finite in the limit $m_Q \to \infty$.

The functions
$C_1$ and $C$ come from the loop integration and are defined as:
\begin{equation}
\int \frac{d^4}{(2\pi)^4} \frac{q_\alpha q_\beta}{(q^2 -m^2)(v \cdot q -\Delta)
(v \cdot q - \Delta')}=
\frac{i}{16\pi^2} \left( C_1(\Delta,\Delta',m) g_{\alpha\beta}
+ C_2(\Delta,\Delta',m)v_\alpha v_\beta \right)
\end{equation}
and
\begin{equation}
C(\Delta,\Delta',m) = C_1(\Delta,\Delta',m) + C_2 (\Delta,\Delta',m) \;.
\end{equation}
Performing the integration we obtain:
\begin{eqnarray}
C_1(\Delta,\Delta',m) & = & \frac{m^3}{9 (\Delta - \Delta')} \left[
H_1 (\frac{\Delta}{m},m ) - H_1(\frac{\Delta'}{m},m) \right]\\
C_(\Delta,\Delta',m) & = & \frac{2 m^3}{9(\Delta - \Delta')} \left[
H (\frac{\Delta}{m},m ) - H(\frac{\Delta'}{m},m) \right]
\end{eqnarray}
where
\begin{eqnarray}
H_1 (x,m) & = & -12 x + 10 x^3 + (9 x - 6 x^3)
\log(\frac{m^2}{\mu^2}) + \nonumber \\
& - & 12 (x^2 - 1)^{3/2} \log (x + \sqrt{x^2 - 1}) \\
H (x,m) & = & - 9 x^3 + (9 x^3 - \frac{9}{2} x ) \log(\frac{m^2}{\mu^2})+
\nonumber \\
& + &  18 x^2 \sqrt{x^2 - 1} \log (x + \sqrt{x^2 - 1}) \; .
\end{eqnarray}
The previous results are obtained in a renormalization scheme such that
$\frac{2}{\epsilon} + \log (4 \pi) -\gamma_E + 1 =0$.

For $\Delta =\Delta'$, as for the wave-function
renormalization factors, we find
\begin{eqnarray}
C_1(\Delta,\Delta,m)= \frac{m^2}{9}H_1'(\frac{\Delta}{m},m) \\
C(\Delta,\Delta,m)= \frac{2 m^2}{9} H' (\frac{\Delta}{m},m)
\end{eqnarray}
where $H'(x,m) = \frac{d H(x,m)}{d x}$.

The diagram of fig. 3 is linear in $h$ (the analogous proportional to $g$
vanishes), and proportional to ${\hat F}^+$, defined in (\ref{coup}).

Combining all the diagrams (fig. 2, 3 and 4 ) we obtain:

\begin{eqnarray}
 f_{D} & =& \frac{{\hat F}}{\sqrt{M_D}} \left[ 1 -
 \frac{1}{32\pi^2 f_\pi^2} \left[\frac{3}{2} m_\pi^2
\log (\frac{m_\pi^2}{\mu^2}) +
 m_K^2 \log (\frac{m_K^2}{\mu^2})+ \frac{1}{6} m_\eta^2
\log (\frac{m_\eta^2}{\mu^2})\right] \right.\nonumber \\
 & - &  \frac{3 g_D^2}{32 \pi^2 f_\pi^2} \left[\frac{3}{2}
C_1(\Delta_{D^* D},\Delta_{D^* D},m_\pi ) +
C_1(\Delta_{D^*_s D},\Delta_{D^*_s D},m_K ) +
 \frac{1}{6} C_1(\Delta_{D^* D},\Delta_{D^* D},m_\eta ) \right] \nonumber \\
& + &  \frac{ h_D^2}{32 \pi^2 f_\pi^2} \left[
\frac{3}{2} C(\Delta_{P_{0} P},\Delta_{P_{0} P},m_\pi ) +
 C(\Delta_{P_{0s} P},\Delta_{P_{0 s} P},m_K ) + \frac{1}{6}
C(\Delta_{P_0 P},\Delta_{P_0 P},m_\eta ) \right] \nonumber \\
& + & \left.\frac{{\hat F}^+}{{\hat F}} \frac{h_D}{16\pi^2 f_\pi^2} \left[
\frac{3}{2}
C(\Delta_{P_0 P},0,m_\pi ) +
C(\Delta_{P_{0s} P},0,m_K ) + \frac{1}{6}
C(\Delta_{P_0 P},0,m_\eta )\right] \right]
\label{fd}
\end{eqnarray}

\begin{eqnarray}
 f_{D_s} & =& \frac{{\hat F}}{\sqrt{M_D}} \left[ 1 -
 \frac{1}{32\pi^2 f_\pi^2} \left[
2 m_K^2 \log (\frac{m_K^2}{\mu^2})+ \frac{2}{3} m_\eta^2
\log (\frac{m_\eta^2}{\mu^2})\right] \right.\nonumber \\
 & - &  \frac{3 g_D^2}{32 \pi^2 f_\pi^2} \left[
2 C_1(\Delta_{D^* D_s},\Delta_{D^* D_s},m_K ) +
 \frac{2}{3} C_1(\Delta_{D^* D},\Delta_{D^* D},m_\eta ) \right] \nonumber \\
& + &  \frac{ h_D^2}{32 \pi^2 f_\pi^2} \left[
2 C(\Delta_{P_{0} P_s},\Delta_{P_{0} P_s},m_K ) + \frac{2}{3}
C(\Delta_{P_0 P},\Delta_{P_0 P},m_\eta ) \right] \nonumber \\
& + & \left.\frac{{\hat F}^+}{{\hat F}} \frac{h_D}{16\pi^2 f_\pi^2} \left[
2 C(\Delta_{P_{0} P_s},0,m_K ) + \frac{2}{3}
C(\Delta_{P_0 P},0,m_\eta )\right] \right] \; .
\label{fds}
\end{eqnarray}

{}From the previous formulae, using $\Delta_{P_0 P} = 0.5 \; GeV$, $\mu = 1$,
${\hat F}^+ = 0.46 \; GeV^{3/2}$ and ${\hat F} = 0.30 \; GeV^{3/2}$,
one gets numerically:
\begin{eqnarray}
f_D & = & \frac{{\hat F}}{\sqrt{M_D}} \left( 1 + 0.09 +
0.11 g^2 -0.33 h^2 -1.00 h \right)
\label{fdnum}
\\
f_{D_s}& = &\frac{{\hat F}}{\sqrt{M_D}} \left( 1 + 0.17 + 0.82 g^2 -
0.66 h^2 - 1.15 h \right) \; .
\label{fdsnum}
\end{eqnarray}
In the previous formulae we have kept only the leading order in the $1/m_Q$,
i.e. we have put $\Delta_{D^*D}=0$ in (\ref{fd}) and (\ref{fds}): therefore
we have to use for the couplings $g_D$ and $h_D$ the asymptotic values
$g$ and $h$ respectively.

The value of $g$ has been computed with QCD sum rules in \cite{g,ruckl,grozin},
giving a result in the range $0.2 - 0.4$.
{}From (\ref{fdnum}) and (\ref{fdsnum}), using $g=0.3$ and $h=-0.5$, we get for
the ratio of leptonic decay constants:
\begin{equation}
\frac{f_{D_s}}{f_D} \simeq 1.09 \; .
\label{ratio}
\end{equation}
 Without the contibution of the excited states, we would get
$f_{D_s}/f_D \simeq 1.13$: the contribution of the excited heavy mesons is
 slightly negative. Notice that the term in $h^2$ tends to cancel against
the term linear in $h$; we also observe that its sign is unambiguously fixed
by the sum rule (see Eq.(\ref{fres}), since the relevant
quantity is the ratio $h  {\hat F}^+/\hat F$.

The result (\ref{ratio}) is not very sensitive to the value of
the mass splitting $\Delta_{D_0 D}$. For instance, if we take
$\Delta_{D_0 D}= 0.6 \; GeV$ we find $f_{D_s}/f_{D} = 1.12$, while for
$\Delta_{D_0 D}= 0.4 \; GeV$ one obtains
$f_{D_s}/f_{D} = 1.05$.
We have also checked that (\ref{ratio}) depends weakly on the value of
the renormalization scale $\mu$.

Keeping only the
``log-enhanced'' terms of the form $m^2 \log (m^2/\mu^2)$, the ratio
becomes:
\begin{eqnarray}
\frac{f_{D_s}}{f_D} & = & 1 - \frac{1}{32 \pi^2 f_\pi^2} \left[
m_K^2 \log (\frac{m_K^2}{\mu^2}) +\frac{1}{2}
m_\eta^2 \log (\frac{m_\eta^2}{\mu^2}) -\frac{3}{2}
m_\pi^2 \log (\frac{m_\pi^2}{\mu^2}) \right] \times \nonumber \\
& \times & \left( 1 + 3 g^2 + h^2 + \frac{{\hat F}^+}{{\hat F}} h \right)
\simeq 1.06
\label{ratiolog}
\end{eqnarray}
The previous formula does not contain the parameter $\Delta_{D_0 D}$.
The quoted number, $1.06$, is for $\mu = 1 \; GeV$: putting $h=0$ in
(\ref{ratiolog}), one obtains the results of \cite{grin,goity},
i.e. Eq.(\ref{gg}).
A different approximation has been put forward in
\cite{falk}, where also terms $\Delta^2 \log (\Delta^2/\mu^2)$ are kept, with
numerical results similar to ours.

In conclusion, the contribution of the excited heavy meson states to the
ratio of leptonic decay constants $f_{D_s}/f_{D}$ is small and negative.
The usual estimate of the $SU(3)$ violation, including only the state $D$ and
$D^*$, is not destabilized when including excited states, at least in this
case. We stress that this result strongly depends on the sign of $h$: had it
been positive, the final result would have been substantially different.
Therefore we cannot exclude that for other observables the  positive
parity heavy mesons give a significant contribution to the chiral loops.
In any case, we point out that the sign of the product
${\hat F}^+/{\hat F} h$, which enters in the formula (\ref{ratiolog}),
 is unambiguously determined
by the sum rule. We also observe that the sign of $h$
turns out to be negative also in ref. \cite{goity1}, based on a chiral quark
model.

The numerical outcome Eq.(\ref{ratiolog}) for
$f_{D_s}/f_{D}$ agrees with the theoretical results obtained by several
groups by different models, e.g. QCD sum rules \cite{noi1},
lattice QCD \cite{lattice} and potential models \cite{noi2}. As for the
experimental results, we only have the upper bound \cite{pdg94}
\begin{equation}
f_D \le 310 \; MeV \label{ddata}
\end{equation}
from MARK III collaboration \cite{mark3}, and the recent results from three
experiments:
\begin{eqnarray}
{f_{D_s}} \; &=& 232 \pm 45 \pm 20 \pm 48 \;MeV \; \; \; \; (WA75 \,
\cite{wa75}) \nonumber \\
{f_{D_s}} \; &=& 344 \pm 37 \pm 52 \pm 42 \;MeV\; \; \; \; (CLEO \,
\cite{cleo}) \nonumber \\
{f_{D_s}} \; &=& 430^{+150}_{-130} \pm 40 \;MeV \; \hskip 1.2cm \; \; \;
(BES \, \cite{bes}) \; . \label{dsdata}
\end{eqnarray}

Even though the theoretical value obtained for
$f_{D_s}/f_{D}$ is still consistent with data  in
(\ref{ddata}),(\ref{dsdata}), it
should be noticed that the experimental results for
$f_{D_s}$ seem to indicate a value much larger than the theoretical estimates
appeared in the literature \cite{noi1,lattice,noi2}, which might signal a
serious theoretical problem. In any event, better quality data are needed
before any conclusion can be drawn.

\section{Conclusions}

We have computed the strong coupling constant of the positive parity heavy
mesons $G_{B^{**} B \pi}$ and $G_{D^{**} D \pi}$,
by QCD sum rules and
light-cone sum rules.
In the limit
$m_Q \to \infty$, the two methods give compatible results, but the
$1/m_Q$ corrections are significant and unfortunately they are rather
 different in
the two approaches.
We have  applied the results to the calculation of the hadronic
widths of the positive parity $B$ and $D$ states: we have found that the
calculated widths are in any case
 compatible with the recent preliminary LEP data on the orbitally
excited $B$ mesons. Furthermore, we have computed the chiral loop
contributions of these
states to the ratio $f_{D_s}/f_D$, and we have found that the chiral
corrections consist of two sizeable quantities, which are however
opposite in sign, so that the prediction
for this ratio obtained using only the
ground state heavy mesons is not significantly shifted.
This cancellation is likely here to be fortuitous, and
in view of the large value we have found for the coupling constant,
leaves open the possibility that
chiral contributions of  the excited heavy mesons to other
physical observables could instead be important.

\acknowledgments{
We thank  F. Feruglio, M. Lenti, A.Palano and
Zhao-Xi Zhang for useful discussions.}


\newpage
\begin{figure}
\epsfxsize=14cm
\hfil\epsfbox{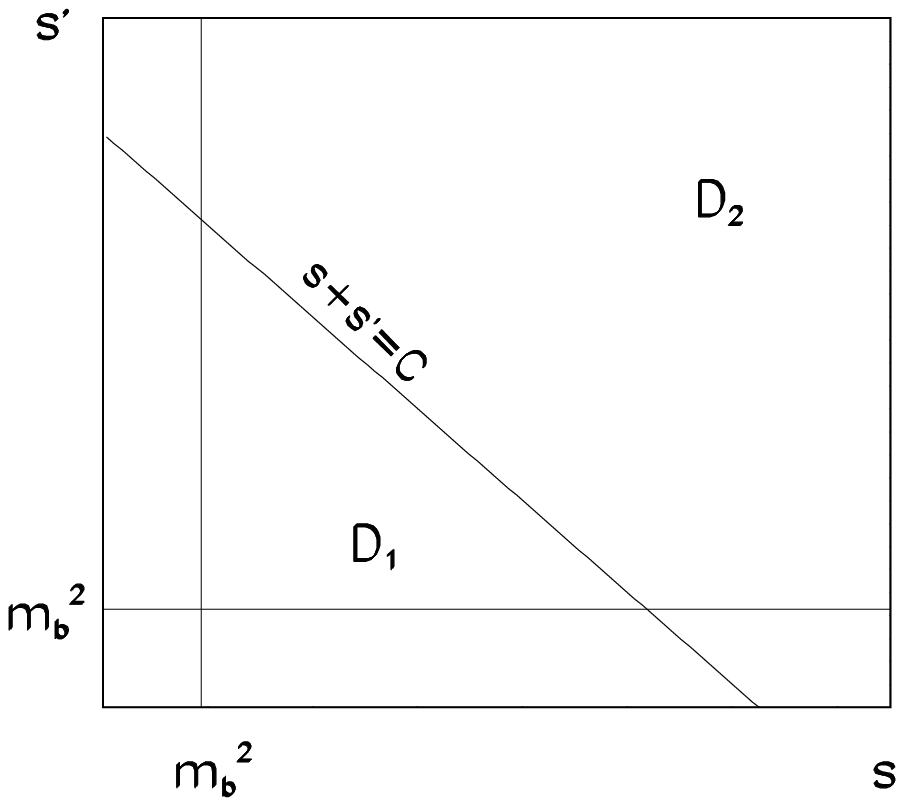}\hfill
\caption{ The integration region of the double dispersive integral
in Eq. (\protect\ref{had}).
In the region $D_1$ only the poles of the $B$, $B^*$ and $B^{**}$ particles are
present, the region $D_2$ includes the hadronic continuum.}
\label{figura}
\end{figure}

\begin{figure}
\epsfxsize=12cm
\hfil\epsfbox{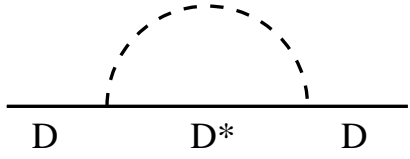}\hfill
\caption{Self energy diagram}
\label{figura2}
\end{figure}

\begin{figure}
\epsfxsize=12cm
\hfil\epsfbox{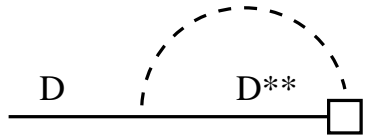}\hfill
\caption{Vertex correction involving positive parity heavy mesons}
\label{figura3}
\end{figure}

\begin{figure}
\epsfxsize=12cm
\hfil\epsfbox{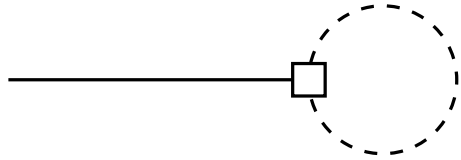}\hfill
\caption{Vertex correction involving only light pseudoscalar mesons}
\label{figura4}
\end{figure}

\end{document}